# Survey Management Web Platform Applied to Morocco Household Survey Panel

El-arbi El-alaouy[1], Khadija Rhoulami[2]

[1]**Faculty of Science, Mohammed V Agdal University**
**Rabat, 4 Av.Ibn Battouta B.P. 1014 RP, Morocco**
*elalaouy.elarbi@gmail.ma*

[2]**Faculty of Science, Mohammed V Agdal University**
**Rabat, 4 Av.Ibn Battouta B.P. 1014 RP, Morocco**
*rhoulamikhadija@yahoo.fr*

**Abstract**
CSPro (Census and Survey Processing System) is a software package used recently in many large scale surveys for data collection. This software often used as desktop software has been used for the first time as a web service for resolving some problems encountered in the first wave of the Morocco Household Panel Survey (MHSP). The article will outline the Survey Management Web Platform that has been developed based on web 2.0 technologies for both integrating the CSPro web service control and centralizing the data files collection from survey fields.

## 1. Introduction

For the first time in Morocco, a household panel system is designed. It involves conducting surveys of a sample of households to be followed at each step wise for a long time. The organizing institution is the National Observatory for Human Development (ONDH) recognized by his monitoring vocation of public policies in the Kingdom. View the scope of this project that covers all regions of Morocco, data collection work and clearance of data were assigned to engineering offices (EO).

CAPI (Computer Assisted Personal Interviewing) is the chosen data collection method [1][2]. Similarly, CSPro is the software that has been selected for data entry [3]. For this purpose, a CSPro application has been developed. For strengthening data privacy and protecting intellectual property of the designed CSPro application, EO has been provided by the CSPro binary data entry application that allows only the data entry (the batch control is not possible with this package).

In addition to this decision, the conventional data transfer using electronic mail from fields to ONDH office and the daily hard work to do for organizing this massive files contained in emails, have caused various problems between ONDH-EO in the first tranche of the initial wave. In parallel of this first tranche, we have thought about designing a platform that on the one hand will integrate as a web service the batch automatic control of data files and on the other hand will manage and centralize the data files transfer from the survey fields.

The benefits of Web 2.0 [4] [5] which tends to displace classical management applications led us to choose in fact to develop a web platform that benefits from all of the Web 2.0 advances. The modeling platform is performed with the Unified Modeling Language UML [6].

In this article, at first we will talk about the MHSP survey and its conduct. Then, we will present the data collection problem followed by the specifications description before starting the web platform modeling in which we present three UML diagrams: use case diagram, sequence diagram and class diagram. We will finally present the Survey Management Web Platform screens before concluding.

## 2. Context of survey

2.1 Presentation of the survey

The MHSP survey has as main objective to realize the monitoring and analysis of the dynamics of human development in Morocco. It cover the main aspects and dimensions of human development (demography, education, health, employment, housing and living conditions, participation, exclusion, subjective poverty, expenditure and consumption, income, etc.) and observe the units "house" and "household" and follow the unit

"household member" in time. The sample of households, followed each year for a long period, consists of 8,000 households selected to represent the country. It comes from the master sample elaborated by the Department of Statistics under the High Commissariat of Plan (HCP).

2.2 Conduct of the survey

The survey will be conducted in all regions of Morocco and affect both the urban and rural areas. The Data collection work and clearance were assigned to engineering offices. They share the national territories, each engineering office is assigned a list of secondary units (U.S) target of survey. A U.S is the smallest elementary unit that is geographically limited. The choice of households to be surveyed at each of the U.S is done, according to a random selection of 2O households from the set of all households in the U.S.

2.3 Problem of data collection

Data collection method used is CAPI [1] [2], it was chosen for its feasibility and time savings and accuracy it provides. A CSPro application has been developed, it allows interviewer to write the responses of household members directly onto computers during the interview. The collecting process distinguishes two type of control:

**Immediate control**: integrated-in CSPro application, it allows to test the validity and consistency of data immediately at the time of entry. To avoid certain blockages that may be caused by this control, and not to increase the work interview, this control was designed to be soft and slight.**Complete control**: ONDH specialist applies this control to the data files sent by an EO Controller. It generates an error listing to be sent again to the Controller.

At the end of each day, the EO controller gathers data files of his interviewers team, send it by email to the ONDH specialist, the latter first organize files in folders and then applies a complete control, then sends the error listing generated to the EO controller, so thereof can correct files as soon as possible. The following figure shows the typical process of transfer and control.

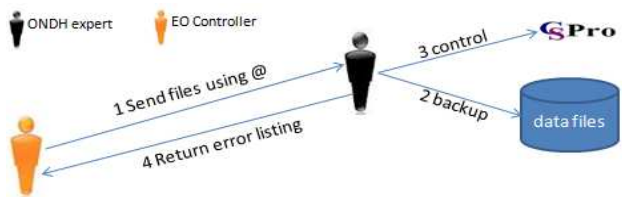

Fig. 1 Process of conventional transfer

There are a total of about fourteen EO controllers, and a team of four interviewers on average per controller, so the number of files to be controlled daily exceeds fifty files, it is an important number to be treated by the ONDH specialist. This central transfer process around the ONDH specialist requires a lot of effort on the part of the latter and EO controllers and causes various problems slowing the survey conduct and increasing its cost.

To solve this problem, we thought to develop a web platform that will allow, on the one hand managing data files transfer, and the other hand offering as web service a complete control of data files automatically without the intervention of ONDH specialist. The following figure illustrates this principle.

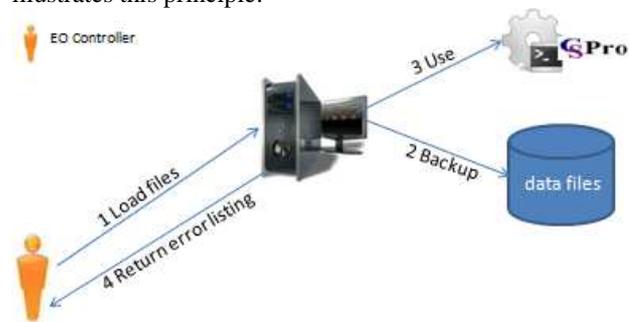

Fig. 2 Principle of the proposed multitasking solution

We define the web platform specifications as follows: Three types of actors were identified (EO Controller, EO Supervisor and ONDH Administrator). The controller performs a daily multiple uploading of data files of his team interviewers. The latter are not interested by the platform. When uploading, a complete control of the data files is done automatically on the server side and an error listing is returned. The U.S data files transferred to the web platform are saved as versions, and organized automatically in a secure folder tree (depending on EO, U.S and upload date). A US version includes all U.S data files transferred on a given day, a U.S. is in its final version when all U.S. households are surveyed. The supervisor follows and monitors the data files collection of controllers and interviewer teams of his EO. The Administrator follows the data files collection of all EO.

Access to the platform is limited. The administrator first creates a supervisor account per EO, Consequently, each supervisor receives his login information. Supervisors authenticate to the platform and start creating the controller accounts of their EO. In the same way, the controllers receive the login information; they can all now authenticate to the platform and use it depending on their profile.

# 3. Modeling

We'll give an overview of the web platform modeling.

## 3.1 Use case diagram

The use case diagram schematizes the functionalities delivered by the platform and expected by users. We have identified four functionalities that are common to all three users: "Authentication" allows user to authenticate; "Edit profile" allows user to change his profile information (name, email and password); "Logout" allows user to disconnect; "Forgot Password" allows user to retrieve the connection information.

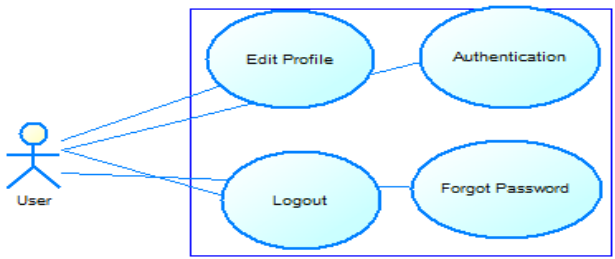

Fig.1  Common use cases

The platform provides the administrator with two fundamental functionalities: "Create supervisor" to create supervisor accounts; "Pursuit Report" to get a pursuit report on the secondary units of all EO, this feature can be optionally achieved by the "U.S control" use case.

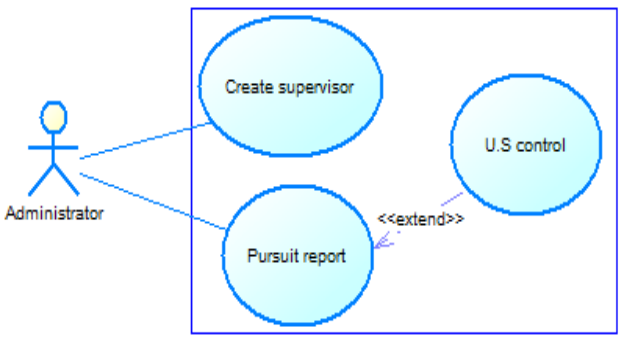

Fig. 2  Use case diagram of administrator

The platform provides the supervisor with two functionalities: "Create Controller" to create controller accounts; "Pursuit Report" to get a pursuit report on the secondary units of its EO, this feature can be optionally accomplished by the "U.S control" use case.

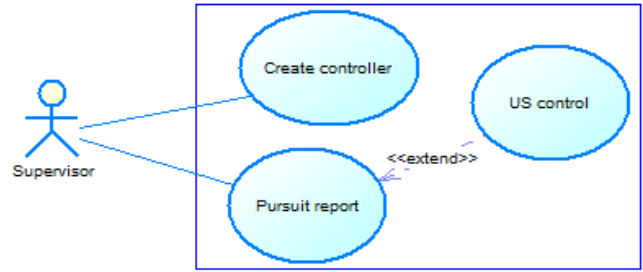

Fig.3  Use case diagram of supervisor

The platform provides the controller with two fundamental functionalities: "Uploading" to upload data files, this feature is automatically accomplished by the "U.S control" use case to apply control on the loaded data files; "Pursuit Report" to get a pursuit report on the secondary units under its responsibility, this latter can be optionally accomplished by the "U.S control" use case.

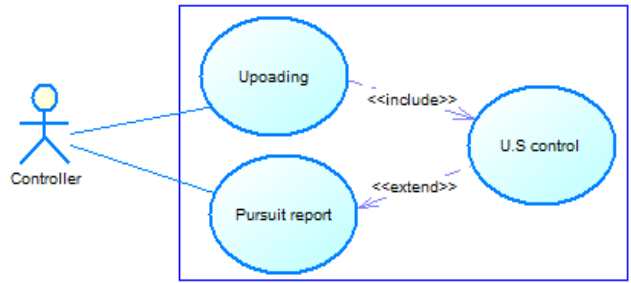

Fig.4  Use case diagram of controller

## 3.2 Sequence diagram

The sequence diagram documents use cases and show the interaction made between user and platform to complete a specific use case. We will present sequence diagrams in the same order we have exposed the use case diagrams:

The user authenticates by typing a username and password and his user category; the platform creates a session and redirects the user to his appropriate space. The following figure describes the authentication functionality.

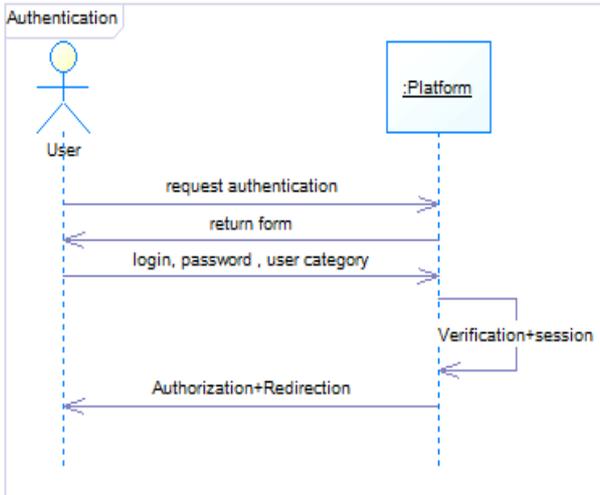

Fig.5 Sequence diagram of authentication

The user requests to logout; the platform deletes his session and redirects him to the default page.

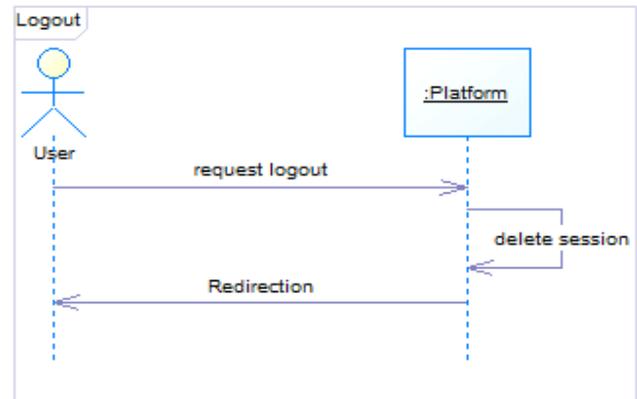

Fig.7 Sequence diagram of logout

The user asks to edit his profile; the platform sends him back the interface; the user enters the new coordinates (name, email and password); the platform checks and edits the information then returns a confirmation message.

The user forgets his password and asks for new login information.

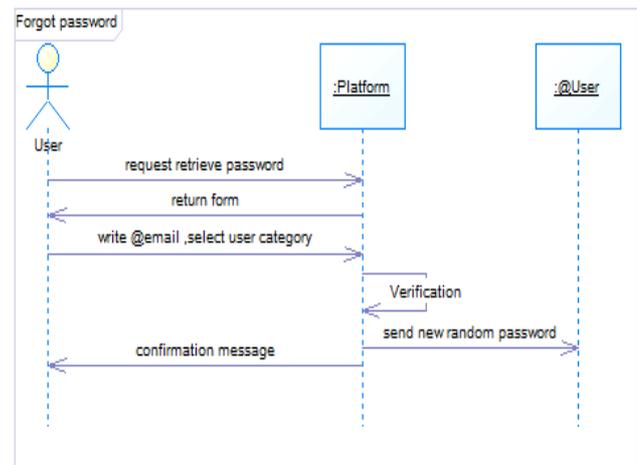

Fig.8 Sequence diagram of forgot password

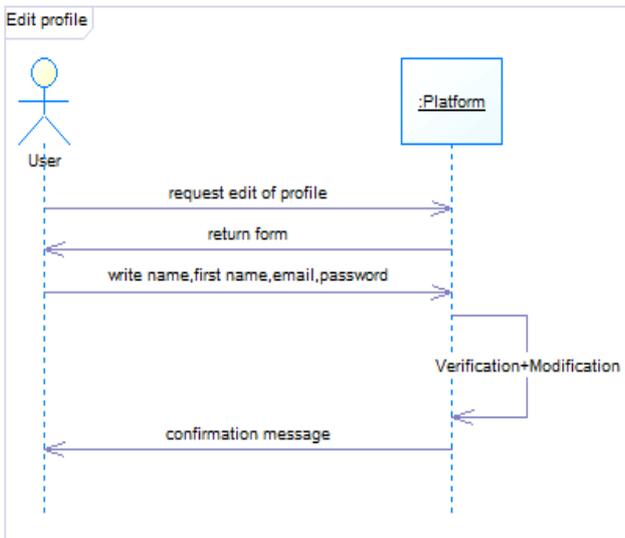

Fig.6 Sequence diagram to edit profile

The Administrator requests the creation of a supervisor account; the platform returns him an interface; the administrator enters the coordinates (name, email, EO); the platform creates the account, and then sends login and a random password to supervisor mailbox; finally, a confirmation message is returned to the administrator.

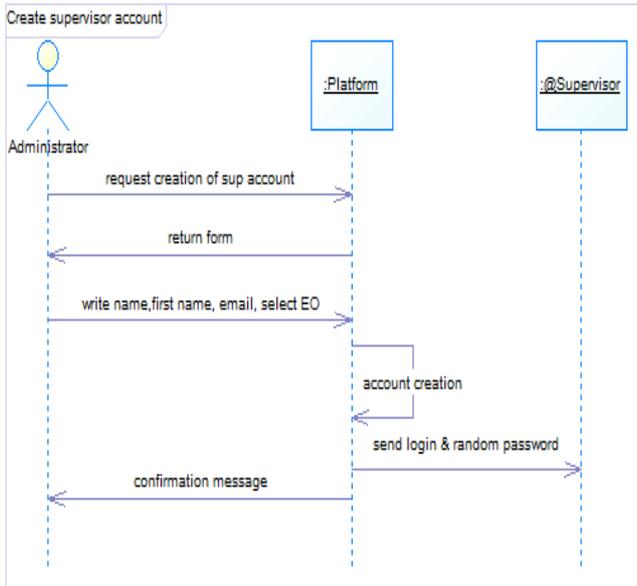

Fig.9 Sequence diagram of creating supervisor account

To generate the pursuit report, the user specifies the search parameters. This pursuit report can give an update and accurate vision of survey work carried out on the fields, it differs depending on the user profile. The controller pursuit report displays only the secondary units under his responsibility. The supervisor pursuit report displays all secondary units of its EO. While the administrator pursuit report displays all secondary units of the various EO.

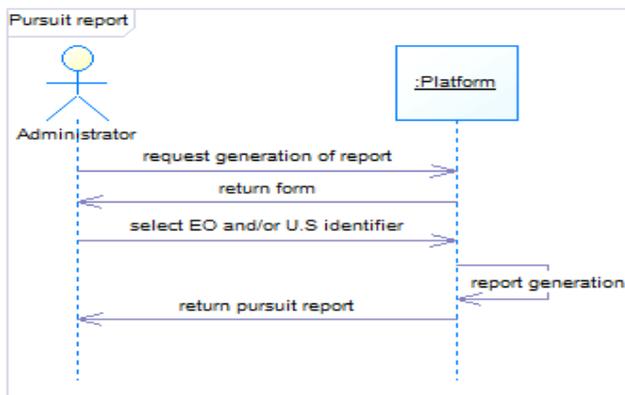

Fig.10 Sequence diagram of pursuit report

The pursuit report can optionally lead to control any secondary unit. This control allows testing the validity and consistency of U.S data and eventually generating an error listing.

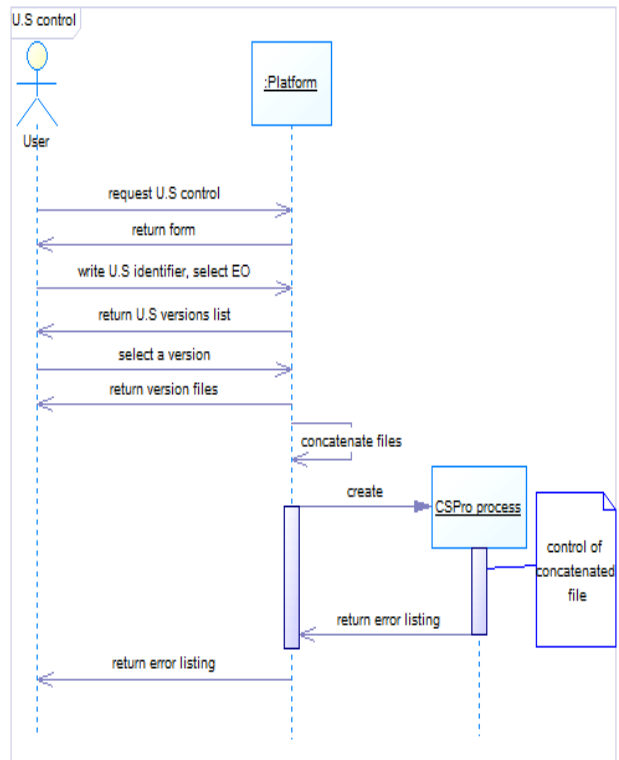

Fig.11 Sequence diagram of U.S control

The supervisor requests the creation of a controller account; the platform returns him an interface; the supervisor enters the code identifying the controller; the platform verifies it, and then returns the name and surname of the controller; the supervisor enters the email address of the controller; the account is created, and then login and a random password are sent to the controller mailbox; finally, a confirmation message is returned to the supervisor. See the figure below.

The controller asks to upload data files; the platform returns him the interface; the controller entry U.S identity, version type, and selects files to load; the platform check, load and save these files in their appropriate folder structure; finally, the platform returns a confirmation message and an error listing.

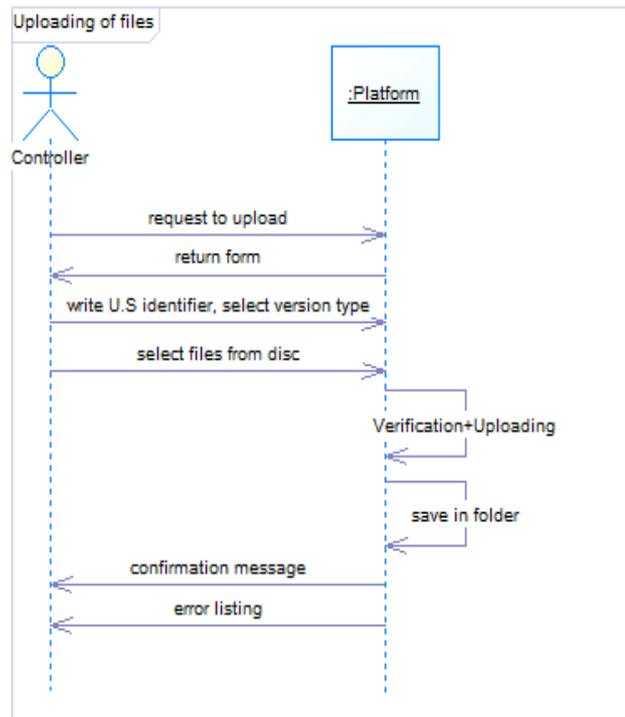

Fig.13 Sequence diagram of uploading

### 3.3 Class diagram

The attempt to abstract the platform domain and separate its elementary entities with similar characteristics has led us to design the class diagram as it is described below.

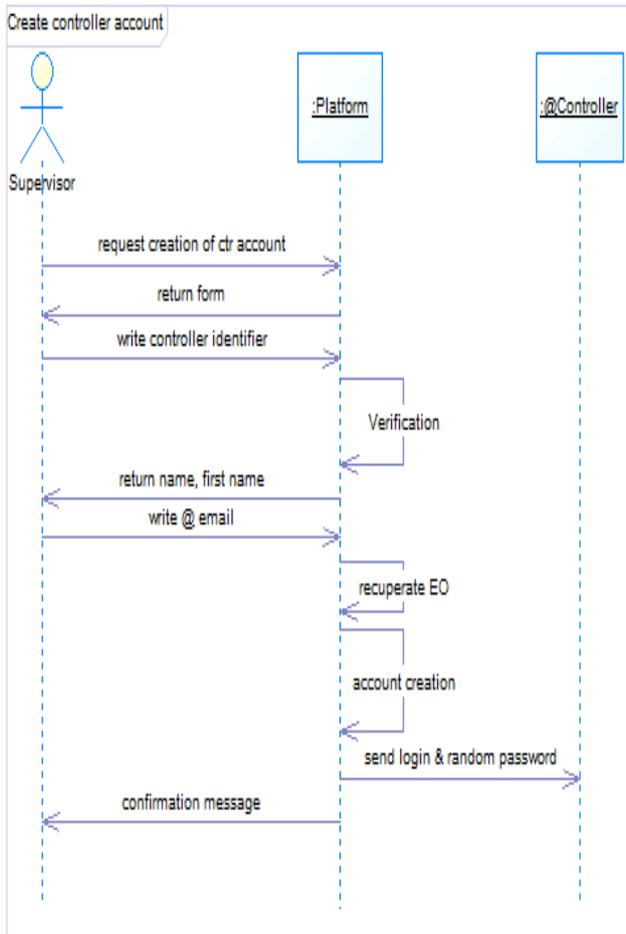

Fig.12 Sequence diagram of creating controller account

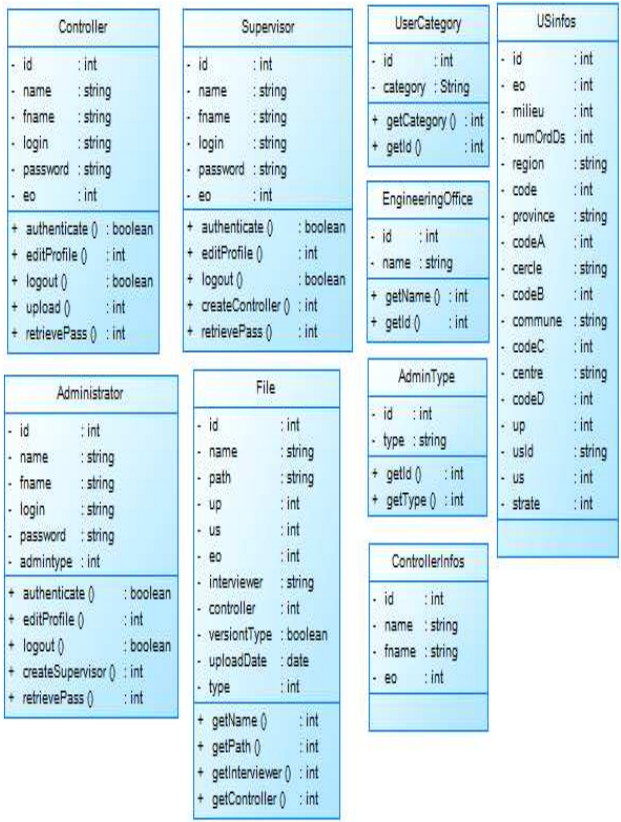

Fig.14 Platform class diagram

## 4. Presentation of the platform

Throughout this section, we will present briefly the main functionalities of the platform. This latter have been developed for French users. Details will not be displayed such as message confirmation, box, mails etc. The platform distinguishes three user profiles.

The default page of the platform consists of a horizontal menu bar (Accueil, Espace réservé, Contact, A-propos), a "Espace réservé" block and a "Forgot password" block.

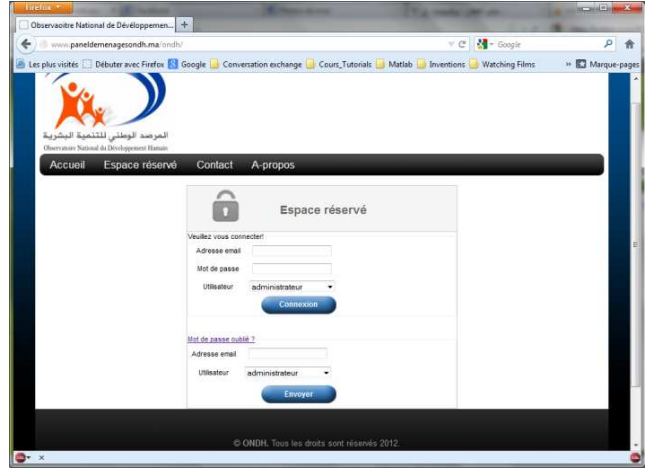

Fig.15 Platform's default page

Access to the platform is restricted. The following figure shows the interface where to authenticate.

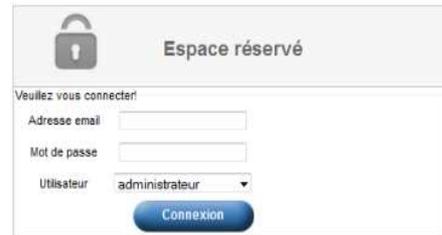

Fig.16 Authentication interface

The reserved space for each user profile consists of a main menu on the left (1), a block at the top right (2) and a block (3) containing the interface that allows the user to perform the requested functionality. Let us start with the dministrator profile, the figure below shows his space.

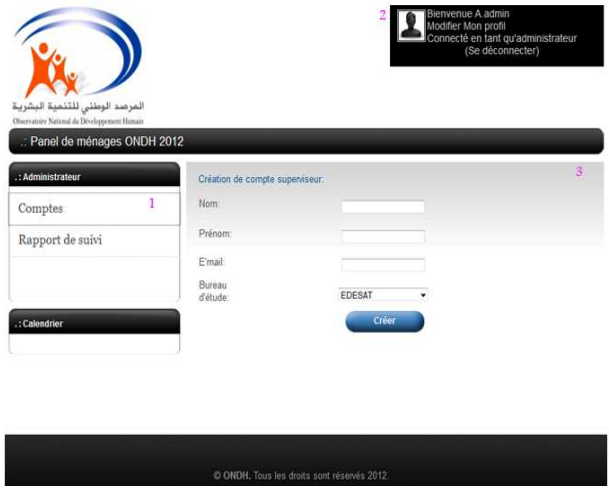

Fig.17 Administrator space

The block (3) is a dynamic block in which the interface responsible for accomplishing the functionality required by the administrator is loaded. The interface of creating the supervisor account is loaded by default in the block (3). It corresponds to the item "Comptes" in the menu (1).

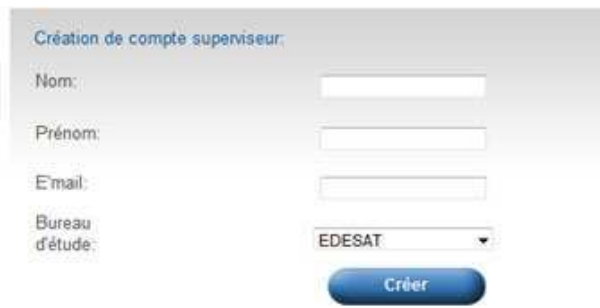

Fig.18 Interface of creating the supervisor account

To generate the pursuit report (see Figure 20), the administrator sets the search parameters in the following interface accessible from the item "Rapport de suivi".

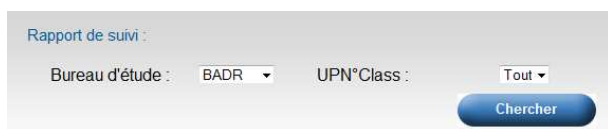

Fig.19 Interface of generating the pursuit report

Fig.20 Pursuit report

To control a secondary unit, the administrator clicks on the "+" button in the intersection of the column "Contrôler" and line of the chosen secondary unit.

Fig.21 Horizontal scroll of pursuit report

Once clicked, a new interface containing the latest version of secondary unit is displayed.

Fig.22 Interface of U.S control

The administrator by clicking the button "Contrôler", he announces a complex control process: The files are concatenated into a new file supplied as a parameter to a function that calls a CSPro core stored at the server side to control the concatenated file and return the generated error listing (Figure 23).

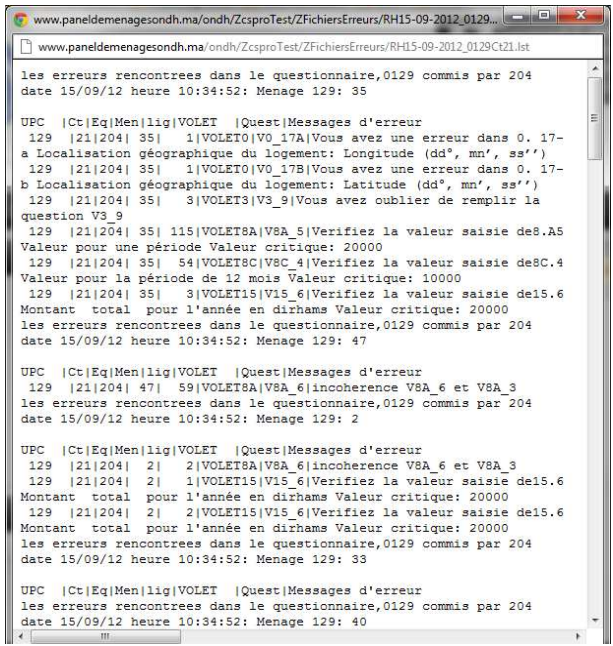

Fig.23 Generated error listing

To edit his profile, the administrator edits his coordinates in the interface accessible from the item "Modifier Profil" in the block (2) at the top right

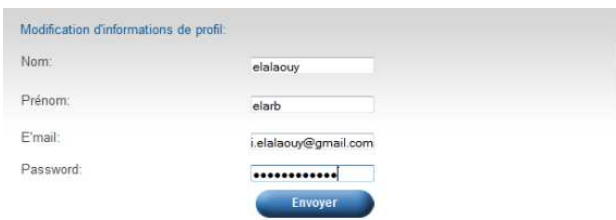

Fig.24 Interface of edit profile

We turn now to the two remaining users. Their menus (1) are as follows.

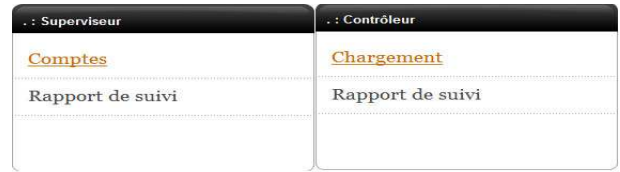

Fig.25 Menus

For the supervisor, he can create a controller account, for that he uses the following interface accessible from the item "Comptes".

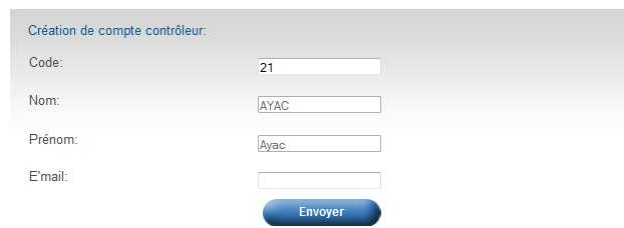

Fig.26 Interface of creating the controller account

The controller when to him, for uploading a multiple data files of a secondary unit, he uses the following interface accessible from the item "Chargement".

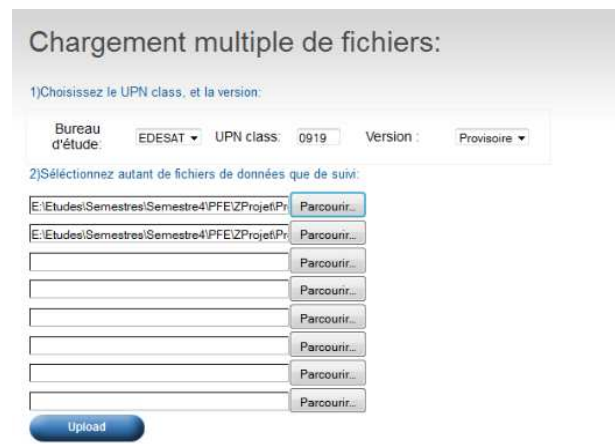

Fig.27 Interface of uploading data files

The item "Rapport de suivi" of the menu (1) of the supervisor and the controller resembles to that of administrator, it allows getting access to the interface of generating pursuit report (Figure 19 and 20). However, this report differs depending on user privileges. The supervisor pursuit report gives an overview of all secondary units of its EO, while the controller pursuit report does a view on the secondary units subject to his responsibility.

## 5. Conclusion

The control batch of CSPro package, used conventionally in many surveys for testing consistency and validity of data files, has been used for the first time, as an online web service control.

This service that has been integrated in the Survey Management Web Platform, allows therefore an automatic control of data files without having to wait to get response from ONDH expert. Furthermore, the platform manages transfer and storage of data files so that users can use daily the web platform to upload data files from their own session wherever they were. With the concept of data files version and automatic folder organization, data files of a given day will not be confused with files of another. As well, files are not missed because they are archived.

The Survey Management Web Platform designed based on web 2.0 technologies, had been used practically by administrator, supervisors and teams of interviewers and gained their satisfaction because of design, simplicities and functionalities it provides. The efficiency of this platform has been proved with the MHSP survey project that covers all regions of Morocco. In the future, we think hopefully to exploit benefits of this platform over a wide range of survey projects.

**El-alaouy El-arbi** Master degree in Computer Science in 2012, PhD student at LRIT laboratory, Mohammed V Agdal University, Faculty of Science, Rabat, Morocco. Ongoing Research interests: residential mobility simulation based on household surveys.

**Rhoulami Khadija** Professor of Computer Science at University Mohammed-V Agdal Rabat, member of Laboratory of Research in Informatics and Telecommunication (LRIT).
Expert and Consultant of surveys Data Processing in ONDH. Ongoing Research interests: modeling and simulating mobility.